\documentclass[
    amsmath,
    amssymb,
    prb,
    reprint,
    superscriptaddress
    ]{revtex4-2}

\usepackage{graphicx}
\usepackage{dcolumn}
\usepackage{bm}
\usepackage{color}
\usepackage{soul}
\begin{document}

\title{Optical-power-dependent splitting of magnetic resonance in nitrogen-vacancy centers in diamond}

\author{Shuji Ito}
\affiliation{Department of Physics, The University of Tokyo, Bunkyo-ku, Tokyo, 113-0033, Japan}
\author{Moeta Tsukamoto}
\affiliation{Department of Physics, The University of Tokyo, Bunkyo-ku, Tokyo, 113-0033, Japan}
\author{Kensuke Ogawa}
\affiliation{Department of Physics, The University of Tokyo, Bunkyo-ku, Tokyo, 113-0033, Japan}
\author{Tokuyuki Teraji}
\affiliation{National Institute for Materials Science, Tsukuba, Ibaraki 305-0044, Japan}
\author{Kento Sasaki}
\affiliation{Department of Physics, The University of Tokyo, Bunkyo-ku, Tokyo, 113-0033, Japan}
\author{Kensuke Kobayashi}
\affiliation{Department of Physics, The University of Tokyo, Bunkyo-ku, Tokyo, 113-0033, Japan}
\affiliation{Institute for Physics of Intelligence, The University of Tokyo, Bunkyo-ku, Tokyo 113-0033, Japan}
\affiliation{Trans-scale Quantum Science Institute, The University of Tokyo, Bunkyo-ku, Tokyo 113-0033, Japan}

\begin{abstract}
Nitrogen-vacancy (NV) centers in diamonds are a powerful tool for accurate magnetic field measurements. 
The key is precisely estimating the field-dependent splitting width of the optically detected magnetic resonance (ODMR) spectra of the NV centers.
In this study, we investigate the optical power dependence of the ODMR spectra using NV ensemble in nanodiamonds (NDs) and a single-crystal bulk diamond.
We find that the splitting width exponentially decays and is saturated as the optical power increases.
Comparison between NDs and a bulk sample shows that while the decay amplitude is sample-dependent, the optical power at which the decay saturates is almost sample-independent.
We propose that this unexpected phenomenon is an intrinsic property of the NV center due to non-axisymmetry deformation or impurities.
Our finding indicates that diamonds with less deformation are advantageous for accurate magnetic field measurements. 
\end{abstract}
\maketitle

\section{Introduction}
A nitrogen-vacancy (NV) center in a diamond is a defect where a nitrogen atom replaces a carbon atom in the lattice with a vacancy at its neighboring site.
The NV center has an electron spin $S=1$, and its peculiar spin-dependent optical transitions enable the optical initialization and readout of the ground-state spin.
This property has been applied to the quantum sensing of local magnetic fields~\cite{MazeNature2008,DegenAPL2008,BalasubramanianNature2008,Taylor2008,SchirhaglARPC2014,Rondin2014,Levine2019,Barry2020} and temperature~\cite{AcostaPRL2010,NeumannNL2013,ToyliPNAS2013}.
Researchers have applied the technique to measure various physical properties, such as observing the electron flow in graphene~\cite{TetienneSciAdv2017,ku2020} and the stray fields from magnetic domain walls of a single-crystal antiferromagnet $\mathrm{Cr_{2}O_{3}}$~\cite{hedrich2021}. 
The basis for these achievements is the ability to accurately measure local magnetic fields on the order of $\mu$T using NV centers.

Optically detected magnetic resonance (ODMR) is a typical and basic measurement technique for quantum sensing using NV centers.
This technique measures the microwave (MW) frequency dependence of the photoluminescence (PL) intensity (red) when the NV centers are continuously irradiated with an excitation light (green) and MW.
The ODMR spectrum presents a magnetic resonance signal between the ground state spin levels $m_S=0$ and $m_S=\pm1$.
The resonance frequency splits against the magnetic field due to the Zeeman effect~\cite{MazeNature2008,DegenAPL2008,BalasubramanianNature2008,Taylor2008,SchirhaglARPC2014,Rondin2014,Levine2019,Barry2020} and shifts in the same direction against temperature change~\cite{AcostaPRL2010,NeumannNL2013,ToyliPNAS2013}.
In addition, the splitting of the resonance frequency is affected by crystal strain~\cite{FoyAPMI2020}, electric field~\cite{VanOort1990,Dolde2011}, and hyperfine interactions~\cite{felton2009}.
Therefore, it is essential for accurate sensing to estimate the splitting width purely due to the magnetic field from the ODMR spectra.
Commonly used diamond samples are single-crystal bulk diamonds and nanodiamonds (NDs) with grain sizes ranging from tens to hundreds of nanometers~\cite{igarashi2012,fu2007}.
Depending on whether the diamond is a bulk crystal or nanoparticles, there are variations in crystal strains, impurity density, and crystal orientation.

The ODMR spectra of NV centers vary with the excitation light power.
For example, the contrast and linewidth vary with the degree of initialization and spin relaxation associated with optical excitation \cite{dreau2011,acosta2013}.
These dependencies only affect sensitivity but not accuracy.

Recently, however, it was reported that the ODMR spectra of NV centers in NDs at low magnetic fields change with the optical power, degrading the accuracy of temperature measurements~\cite{fujiwara2020}.
They found that a change in the ODMR splitting up to $2.8~\mathrm{MHz}$ (equivalent to Zeeman splitting for 50~$\mu$T) occurred depending on the optical power.
This unexpected observation directly affects the accuracy of the conversion of the ODMR splitting to magnetic field, which is a critical issue in achieving the $\mu$T-order magnetic field measurements necessary for the physical properties measurements.
In particular, in wide-field imaging of magnetic field and temperature using a CMOS camera and NV ensembles~\cite{FoyAPMI2020,ScholtenJAP2021,TsukamotoAPL2021,Tsukamoto2022}, inhomogeneity of the optical power within the field of view could result in degradation of the measurement of the magnetic field and temperature distributions.
Thus, it is crucial to investigate the extent to which this phenomenon is universal for various samples, i.e., bulk diamonds as well as NDs.

In this study, we investigate the dependence of the ODMR splitting on the optical power using several NV ensemble samples.
We first investigate the NV ensembles in NDs with a grain size of 100 nm, the same size as in the previous study~\cite{fujiwara2020}.  
We confirm the reported behavior of the ODMR splitting to decrease with increasing optical power.
In addition, we measure the ODMR spectra over a broader optical power range than in the previous study. 
We thereby find the splitting decays exponentially with the optical power and saturates at a constant value.
We observe similar behavior in NDs with a different grain size of 50 nm. 
We then investigate NV ensembles in a single-crystal bulk diamond with much fewer impurities and strain than NDs and find a weaker but similar behavior.
We prove the irrelevance of magnetic field and temperature on this observation and discuss possible mechanisms to account for this phenomenon. 
Finally, we propose the possibility that repetitive photoionization of impurities averages the local non-axisymmetry environment of NV centers and a systematic method to deal with this phenomenon.

This paper is organized as follows.
Sec.~\ref{sec:exp} describes the experimental setup and defines the optical power in this study.
Sec.~\ref{subsec:randd} reproduces the previous study~\cite{fujiwara2020} using NDs and confirms that the ODMR spectra change with optical power.
Sec.~\ref{sec:bulkdia} shows that a similar phenomenon occurs even in the single-crystal bulk diamond.
Sec.~\ref{subsec:delta} analyzes the dependence of the ODMR splitting on the optical power.
    In Sec.~\ref{subsec:poss}, we discuss the influence of the magnetic field and temperature, possible mechanisms, and implications of the present finding.
Sec.~\ref{sec:conc} presents our conclusions.

\section{Experiments}\label{sec:exp}
Figure 1(a) shows an overview of the experimental setup~\cite{misonou2020}.
All measurements in this study are performed in a confocal system at room temperature.
A green laser with a wavelength of 520~nm (Oxxius, LBX-520-70-CSB-PPA) is applied for initialization and readout of the NV centers.
The intensity of the green laser is adjusted using several fixed neutral density filters as appropriate.
The intensity of the red emission from the NV centers is detected by an avalanche photodiode (APD) after passing through a dichroic mirror, a 514 nm notch filter, a 650 nm long-pass filter, and an 800 nm short-pass filter. 
When measuring NV centers in nanodiamonds, the red emission counts were suppressed using a fixed neutral density filter to match the APD measurement range.
We use a MW antenna for spin manipulation of the NV centers, which is a coplanar waveguide with ground consisting of a 1.6~mm thick PCB substrate and an 18~$\mu$m thick copper foil with a 2~mm width centerline terminated with a 50~$\Omega$ resistor.
The antenna is impedance matched so that no frequency dependence of the MW power at a sample position is present during the measurement.
We confirm that from S11 parameter.
Microwaves are output from a vector signal generator at approximately $-13$~dBm and input to a microwave antenna after passing through an MW amplifier (typ.~$+45$~dB).
In all measurements in this paper, the microwave power is fixed at the above values.

\begin{figure}[tbp] 
\includegraphics[width = \linewidth]{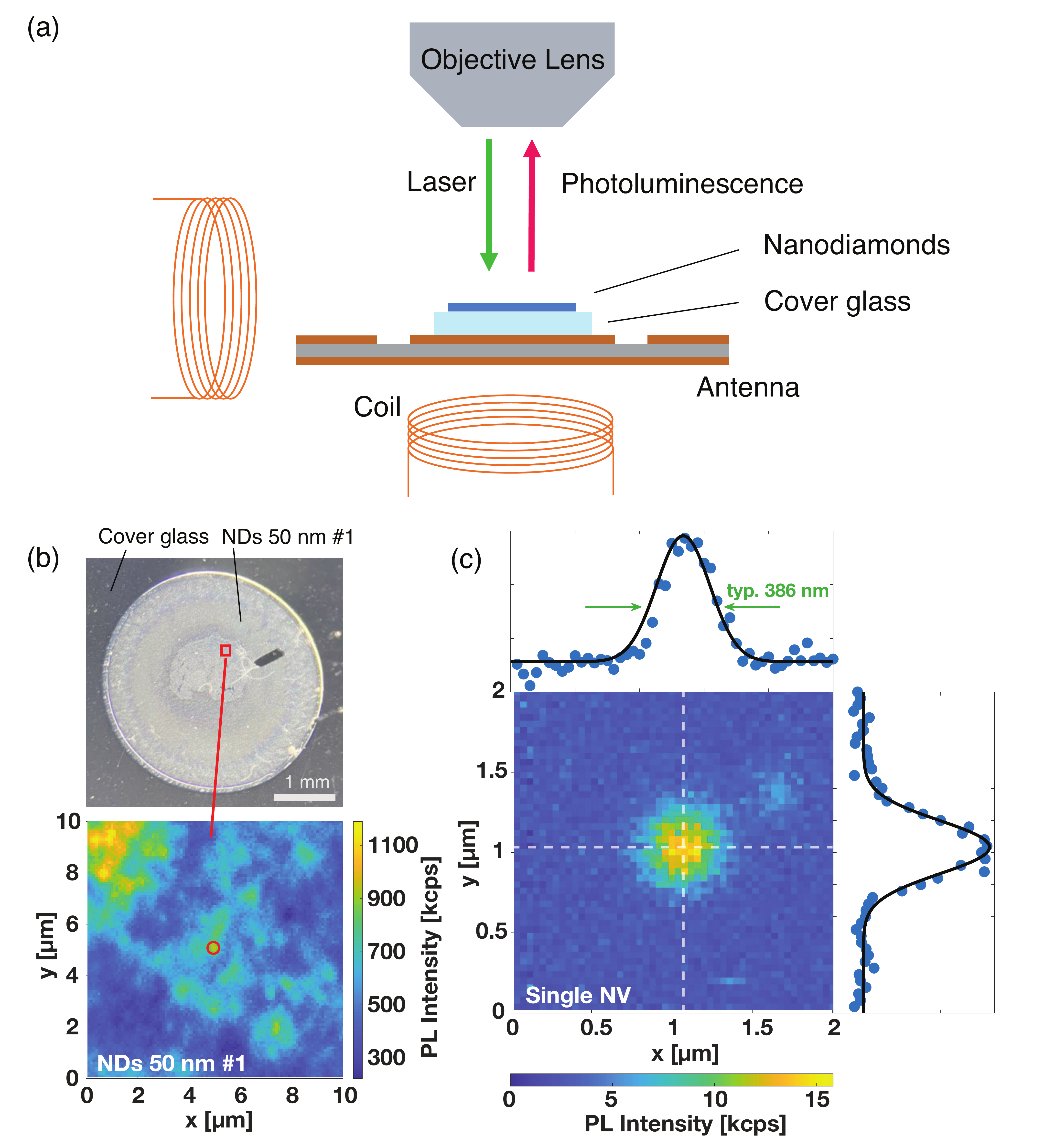}
\caption{\label{fig:1}
(a) Experimental setup.
A wavelength of the green laser is 520~nm.
The MW antenna is a coplanar waveguide with ground.
The two coils beside and beneath the sample stage are for applying the magnetic fields perpendicular and parallel to the optical axis, respectively.
Nanodiamonds spin-coated on the cover glass (\#1 and \#2) are fixed to the antenna with carbon tape and placed on the sample stage as shown here.
In the case of the bulk diamond film (\#3), we fix it directly to the antenna with carbon tape.
(b) The upper figure shows an optical microscope image of a spin-coated ND film (\#1).
The lower figure shows a PL intensity map of the spot surrounded by a red frame in the top image.
The color bar indicates PL intensity in a unit of kilo counts per sec (kcps).
The red circle is the ODMR measurement spot.
(c) Determination of the irradiation area for the optical power calibration.
The color map shows the PL intensity distribution from a single NV center.
The cross sections of the color map are shown in markers in the top and side panels with the results of the 2D Gaussian fitting (solid line).
}
\end{figure}

We use three types of diamond samples, \#1, \#2, and \#3, in the present study: NDs with nominal grain sizes of $\phi50$ nm (\#1) and of $\phi100$ nm (\#2), and NV ensemble in a bulk diamond film (\#3).

The NDs are those commercially available from Ad\'{a}mas Nanotechnologies, NDNV50nmHi10ml for \#1 and NDNV100nm10ml for \#2.
In the measurements of \#1 and \#2, we prepare a ND film [see Fig.~1(b)], which is the NDs spin-coated on a cover glass at 600~rpm~\cite{Tsukamoto2022}. 
The thickness of the ND film made by this method is typically about 200--1000~nm~\cite{Tsukamoto2022,OgawaJPSJ2023}.
The number of NDs in \#1 and \#2 within a laser irradiation area is estimated to be several hundred and more than 20, respectively.
The ND film is fixed to the antenna with carbon tape.
A surface of the ND film is at a height of 0.44 mm above the antenna.

In addition to NDs, this study investigates a bulk diamond film (\#3). 
It was synthesized using a custom-built microwave plasma chemical vapor deposition (MPCVD) system~\cite{TerajiPSSA2015}. 
High-pressure and high-temperature type-Ib (100) single crystalline diamond plates were used as substrates. $^{12}$C concentrated ($>99.95$\%) methane gas was used as a carbon source. 
First, an undoped thick film with a total thickness of $\sim$70~$\mu$m was grown on the substrate by chemical vapor deposition (CVD). 
A $^{15}$N doped CVD layer was then overgrown on the undoped film with a gas ratio of $^{15}$N$/$C of 4000~ppm. 
An expected $^{15}$N concentration is $\sim$10~ppm and a film thickness is $\sim$5~$\mu$m.
This nitrogen density is consistent with the NV's coherence $T_2 = 29~\mu$s obtained by Hahn echo~\cite{Bauch2020}.
We fix \#3 directly to the antenna with carbon tape for the measurement.
A surface of the bulk diamond film is at a height of 0.73 mm above the antenna.
In this study, NV centers spontaneously formed during the MPCVD process are used for characterization.

We perform the present study under three different magnetic fields: a zero field (A), an environmental field (B), and a biased field (C).
We apply magnetic fields for the conditions A and C. 
We use two coils beside and beneath the sample stage to generate magnetic fields perpendicular and parallel to the optical axis, respectively, as shown in Fig.~1(a).
Using a tesla meter (Lake Shore Cryotronics F71), we evaluate the magnetic fields at the sample position as $6.3~\mathrm{\mu T}$, $88.7~\mathrm{\mu T}$, and $196.7~\mathrm{\mu T}$ for the conditions A, B, and C, respectively.

The upper panel of Fig.~1(b) shows an optical microscope image of the spin-coated NDs $\phi50$ nm (\#1). 
The lower panel shows the PL intensity map at the spot surrounded by a red frame in the upper panel.
The color bar indicates PL intensity in a unit of kilo counts per sec (kcps).
The data set for \#1 is obtained using the standard ODMR measurement at the red circle.

As the dependence of the ODMR spectra on the optical power of the excitation light is the central topic in this study, it is important to calibrate the optical power ($P_\mathrm{opt}$). 
We evaluate $P_\mathrm{opt}$ from the green laser intensity and the irradiated area with an accuracy of $10\,\%$.
The green laser intensity is measured between the objective lens and the diamond sample using an optical power meter (Thorlab, Power Meter PM100D, sensor S121C).
The irradiation area is estimated as the spot size of the red luminescence from a single NV center near the surface of a high quality bulk diamond provided by H. Watanabe in AIST, Japan \cite{ohashi2013}.
The spot size is calculated as a circle whose diameter is the full width at half maximum of the intensity distribution.
Figure 1(c) presents an example of the PL intensity map from a single NV center used to determine the spot size, where the diamond surface is defined as the $xy$-plane.
Ten PL intensity maps of a single NV center are fitted by the two-dimensional (2D) Gaussian function, and the obtained average of their full width at half-maximum, $386~\pm~2$~nm, is used as the laser spot diameter.
The cross sections of the experimental data (markers) and the 2D Gaussian fitting (solid line) are shown in the upper side and right side panels of Fig.~1(c).
Both panels show that the fits are consistent with the experimental data.

All the experimental conditions in this study are compiled in Table~\ref{tab:condition}.
NDs $\phi100$ nm \#2' in Table~\ref{tab:condition} indicates the data set obtained at a different location of the same sample as NDs  $\phi100$ nm \#2.
The estimated densities of nitrogen, $[\text{N}]$, and NV center, $[\text{NV}]$, are also given in Table~\ref{tab:condition}. 
We include the previous study (Ref.~\cite{fujiwara2020}) in Table~\ref{tab:condition} in the same cell as 2B as their measurements were carried out in an environmental geomagnetic field ($\sim50~\mathrm{\mu T}$) using NDs $\phi100$~nm supplied by Ad\'{a}mas Nanotechnologies.

\begin{table}
    \centering
    {\tabcolsep = 0.5mm 
    \begin{tabular}{|c||c|c|c|} \hline
        Condition & zero (A) & envir. (B) & biased (C) \\ \hline
        \hline
        \begin{tabular}{c} NDs $\phi50$ nm \#1 \\ $[\text{N}]\sim100$~ppm \\ $[\text{NV}]\sim2$~ppm  \end{tabular} & 1A & - & - \\ 
        \hline
        \begin{tabular}{c} NDs $\phi100$ nm \#2, \#2' \\ $[\text{N}]\sim100$~ppm  \\ $[\text{NV}]\sim3$~ppm  \end{tabular} & \begin{tabular}{c} 2A \\ 2'A  \end{tabular}  & \begin{tabular}{c} 2B \\ Ref.~\cite{fujiwara2020}  \end{tabular}  & 2C \\ 
        \hline
        \begin{tabular}{c} Bulk diamond \#3 \\ $[\text{N}]\sim10$~ppm \\ $[\text{NV}]\sim4$~ppb  \end{tabular} & 3A & - & - \\
        \hline
    \end{tabular}}
    \caption{
    \#1, \#2, and \#3 correspond to the NDs $\phi50$ nm, $\phi100$ nm, and the bulk diamond film, respectively (see text).
    $[\text{N}]$ and $[\text{NV}]$ are given for each.
    \#2' means the data set obtained at a different location of the sample \#2.
    The magnetic field conditions A, B, and C correspond to 
    a zero field ($6.3~\mathrm{\mu T}$), an environmental field ($88.7~\mathrm{\mu T}$), and a biased field ($196.7~\mathrm{\mu T}$), respectively. The experimental condition of Ref.~\cite{fujiwara2020} is classified in 2B in the present study.
    }
    \label{tab:condition}
\end{table}

\section{Results and Discussions}\label{sec:randd}
\subsection{ODMR Spectra of Nanodiamond NVs}\label{subsec:randd}
The upper panel of Fig.~2(a) is the ODMR spectrum as a function of the MW frequency obtained at  $P_\mathrm{opt}=0.55~\mathrm{kW/cm^2}$ shown by markers. 
This result is for 2A (see Table~\ref{tab:condition}).
The vertical axis indicates the PL contrast, namely the normalized contrast of the PL intensities with and without MW.
In this measurement, the swept frequency range is 60 MHz.
The splitting between dips in the ODMR spectrum is due to crystal strain and electric fields that break the axial symmetry of the NV centers. The impacts of such non-axisymmetry deformation were treated in Refs.~\cite{FoyAPMI2020, VanOort1990,Dolde2011}.
Below we call these factors as ``deformation''.

We note that similar observations for the NDs ensemble were reported before, for example, in Fig.~1(d) of Ref.~\cite{FoyAPMI2020}. Their shapes are generally consistent with ours, while the splitting is slightly larger than that in the present study as they applied a magnetic field of 100 $\mathrm{\mu T}$.
Also, similar ODMR spectra obtained in a single ND were reported in Fig.~3(a) of Ref.~\cite{fujiwara2020}.

\begin{figure} 
\includegraphics[width = \linewidth]{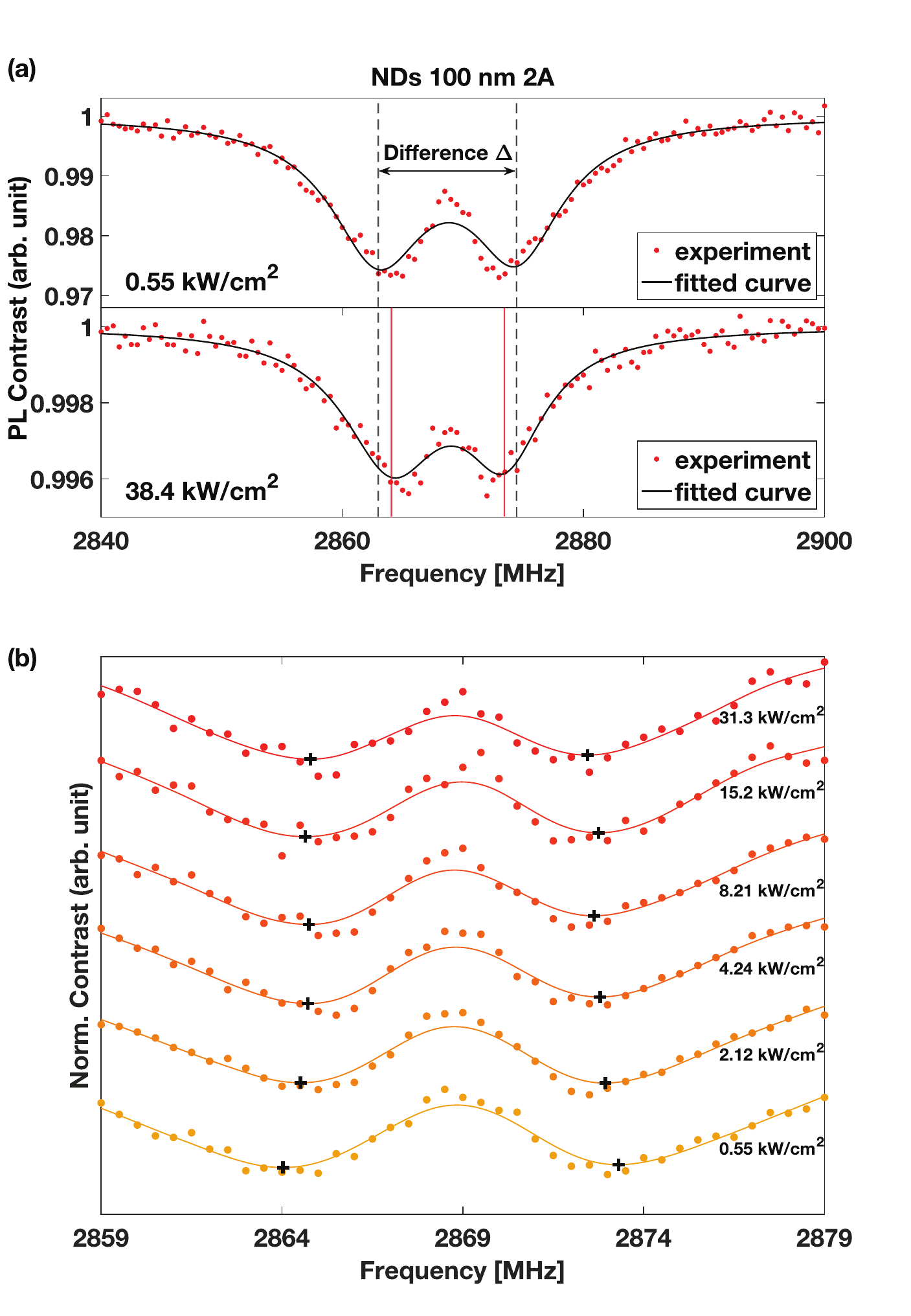}
\caption{\label{fig:2}
ODMR spectra of NDs $\phi100$~nm at zero magnetic fields (the condition 2A).
(a) Spectra measured at $P_\mathrm{opt}=0.55~\mathrm{kW/cm^2}$ in the upper panel and that at $P_\mathrm{opt}=38.4~\mathrm{kW/cm^2}$ in the lower panel are shown in markers.
The solid cuved lines are the results of the fitting.
The dashed and solid vertical lines are drawn at the dip positions deduced from the fitting in the upper and lower panels, respectively.
This clearly indicates that $\Delta$ decreases as $P_\mathrm{opt}$ increases.
(b) ODMR spectra at various $P_\mathrm{opt}$'s.
They are incrementally shifted from bottom to top in the order of $P_\mathrm{opt}=0.55, 2.12, 4.24, 8.21, 15.2$, and $31.3~\mathrm{kW/cm^2}$ as shown by the numbers in the right column.
Markers are experimental data, and solid lines are the spline interpolation curves instead of the fitted ones.
Cross markers ($+$) indicate the dip positions in the interpolated curve.
}
\end{figure}

From now on, we focus on splitting quantitatively based on the values obtained from fitting with a double Lorentzian function. 
This fitting method is meaningful because it is often used for magnetometry using NVs.
We will discuss the validity and limitations of this method later in Sec.~\ref{subsec:poss}.
The solid line in the upper panel of Fig.~2(a) is a fitted curve.
We define the difference in frequencies between the two dip values obtained by this fitting as the difference $\Delta$.
$\Delta$ is $11.5\pm0.2~\mathrm{MHz}$ in this specific case, which is consistent with the literature values of 10--20~MHz for NDs~\cite{FoyAPMI2020,fujiwara2020}.

We measure the ODMR spectra by increasing $P_\mathrm{opt}$ from $0.55~\mathrm{kW/cm^2}$.
The lower panel of Fig.~2(a) shows the spectrum for 2A obtained at $P_\mathrm{opt}=38.4~\mathrm{kW/cm^2}$, which is the maximum optical power used in the present study.
We discuss later that the temperature increase due to laser heating is inconsequential within the present optical power range.
As in the upper panel, the markers show experimental data, and the solid curved line results from a double Lorentzian fitting.
The PL contrast decreases from 2.7\% at $P_\mathrm{opt}=0.55~\mathrm{kW/cm^2}$ to $0.5$\% at $P_\mathrm{opt}=38.4~\mathrm{kW/cm^2}$ because the increase in the optical power enhances the spin initialization rate, i.e., the transition rate from $m_S=\pm1$ to $m_S = 0$.
The spectrum also possesses two dips, but careful inspection reveals a slight change in shape between the upper and lower panels.
The dashed and solid vertical lines show the dip positions obtained by the fitting at $P_\mathrm{opt}=0.55~\mathrm{kW/cm^2}$ and $P_\mathrm{opt}=38.4~\mathrm{kW/cm^2}$, respectively.
$\Delta$ is determined to be $9.4\pm0.3~\mathrm{MHz}$ for $P_\mathrm{opt}=38.4~\mathrm{kW/cm^2}$.
Thus, $\Delta$ decreases with increasing $P_\mathrm{opt}$.

Similar behavior was reported in Fig.~3(a) of Ref.~\cite{fujiwara2020}, suggesting that $\Delta$ of NVs in NDs actually depends on the optical power, which is usually not considered.
In our case, $\Delta$ changes by approximately 2.1~MHz between the two different $P_\mathrm{opt}$.
Significantly, ignoring deformation, this variation corresponds to about 38~$\mu$T according to a magnetic field conversion widely used in the NV research field.
Therefore, this phenomenon can be relevant in applying NVs to magnetic field measurements.

The above finding is not an artifact caused by a double Lorentzian fitting.
To confirm this, Fig.~2(b) presents the ODMR spectra measured at $P_\mathrm{opt}=0.55, 2.12, 4.24, 8.21, 15.2,$ and $31.3~\mathrm{kW/cm^2}$, which are incrementally shifted from bottom to top. 
The markers are the experimental data, where the spline interpolation curves are superposed by the solid lines. 
Since the PL contrast varies depending on $P_\mathrm{opt}$, we appropriately normalize the spectra to focus only on the shape. 
The cross markers ($+$) point to the dip positions in the spline interpolation curves.
Their behavior again supports that the two dips become closer for a larger $P_\mathrm{opt}$.

While we do not show the data, the results of the condition 2'A and the NDs of $\phi50$~nm (1A) are consistent with the results of 2A. Some results are later shown in Figs.~4(d), 4(e), and 4(f). 

\subsection{ODMR Spectra of Bulk Diamond NVs}\label{sec:bulkdia}
We focus on the bulk diamond film \#3 to investigate whether or not the optical power dependence observed in NDs is relevant here.
The upper panel of Fig.~3 presents the ODMR spectrum for the condition 3A obtained at $P_\mathrm{opt}=0.55~\mathrm{kW/cm^2}$.
The horizontal axis range is 10~MHz, much smaller than that in Fig.~2(a).
The obtained spectrum shown by the markers has two sharp dips, as expected for the NVs in bulk diamonds.
As performed for the analysis of NDs, we fit the experimental data with a double Lorentzian function.
We estimate the splitting between the two dips to be $\Delta=3.55\pm0.02~\mathrm{MHz}$, a comparable value to the width of 3.03 MHz due to the hyperfine interaction in $^{15}\mathrm{N}$~\cite{felton2009}.
Presumably, the deformation is much less than $1~\mathrm{MHz}$ because it is buried in this hyperfine splitting.
Thus, the bulk diamond differs from NDs because the hyperfine interaction prevails over the deformation.
In addition, the resonance line width is significantly narrower than in the NDs.
This reflects that the density of impurities, such as nitrogen impurities (P1 centers), which cause the decoherence~\cite{Barry2020}, is low in \#3.
Indeed, the typical nitrogen concentration of a type 1b diamond, the raw material of NDs, is about 100 ppm, whereas the single-crystal diamond in this study is about 10~ppm.

\begin{figure} 
\includegraphics[width = \linewidth]{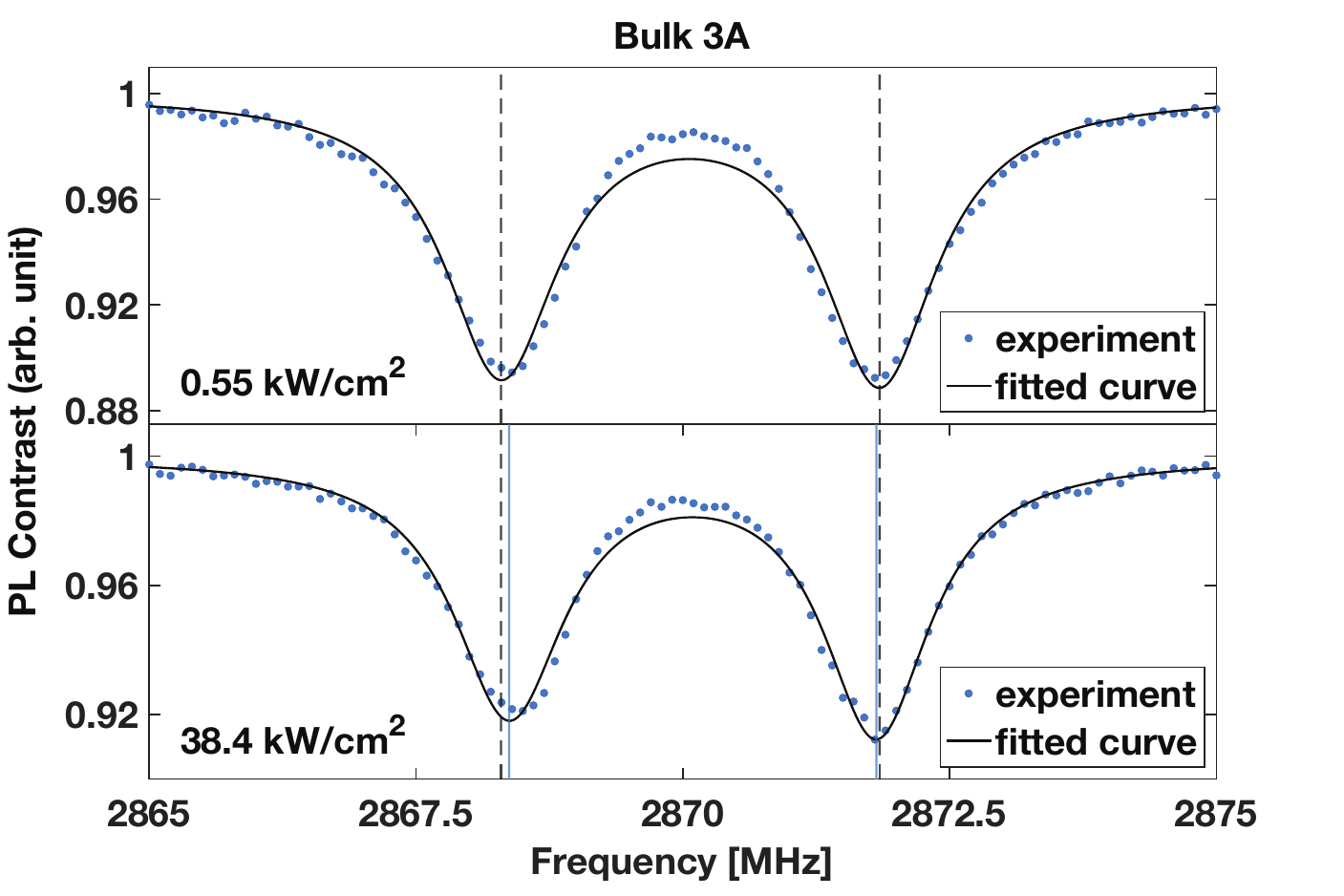}
\caption{\label{fig:3} ODMR spectra of the bulk diamond film \#3 at zero magnetic fields (the condition 3A, see Table~\ref{tab:condition}) measured at $P_\mathrm{opt}=0.55~\mathrm{kW/cm^2}$ in the upper panel and $P_\mathrm{opt}=38.4~\mathrm{kW/cm^2}$ in the lower panel.
The markers are the experimental data with the fitted curve shown by the solid line.
The dashed and solid vertical lines are drawn at the fitted dip values in the upper and lower panels, respectively.
}
\end{figure}

Now, we discuss the ODMR spectra at increased optical powers.
The lower panel in Fig.~3 shows the ODMR spectrum by the markers in the condition 3A obtained at $P_\mathrm{opt}=38.4~\mathrm{kW/cm^2}$.
The markers are experimental data, and the solid curved line results from a double Lorentzian function fitting.
As seen in NDs, the contrast decrease is also due to a larger initialization rate in larger optical power.
In Fig.~3, the dashed and solid vertical lines indicate the dip positions obtained by the fitting at $P_\mathrm{opt}=0.55~\mathrm{kW/cm^2}$ and $P_\mathrm{opt}=38.4~\mathrm{kW/cm^2}$, respectively.
$\Delta$ is now $3.44\pm0.01~\mathrm{MHz}$, smaller than $\Delta=3.55\pm0.02~\mathrm{MHz}$. 
As in the NDs case, $\Delta$ becomes smaller in the larger optical power in the bulk diamond.
Interestingly, the optical power dependence is present even when the $^{15}\mathrm{N}$ hyperfine interaction causes the splitting.
However, the reduction of $\Delta$ in the bulk diamond is much smaller than in NDs.

\subsection{Analysis of Splitting}\label{subsec:delta}
We systematically examine the dependence of $\Delta$ on $P_\mathrm{opt}$.
We start with the condition 2A.
The upward triangle markers in Fig.~4(a) show the experimentally observed $\Delta$ as a function of $P_\mathrm{opt}$ between $0.55~\mathrm{kW/cm^2}$ and $38.4~\mathrm{kW/cm^2}$.
We already showed the results of $\Delta$ at the minimum ($P_\mathrm{opt}=0.55~\mathrm{kW/cm^2}$) and maximum ($P_\mathrm{opt}=38.4~\mathrm{kW/cm^2}$) optical powers in the upper and lower panels in Fig.~2(a), respectively.
Figure~4(a) clearly tells that $\Delta$ monotonously decays with increasing $P_\mathrm{opt}$ and saturates at $P_\mathrm{opt} \gtrsim 15~\mathrm{kW/cm^2}$.

\begin{figure} 
\includegraphics[width = \linewidth]{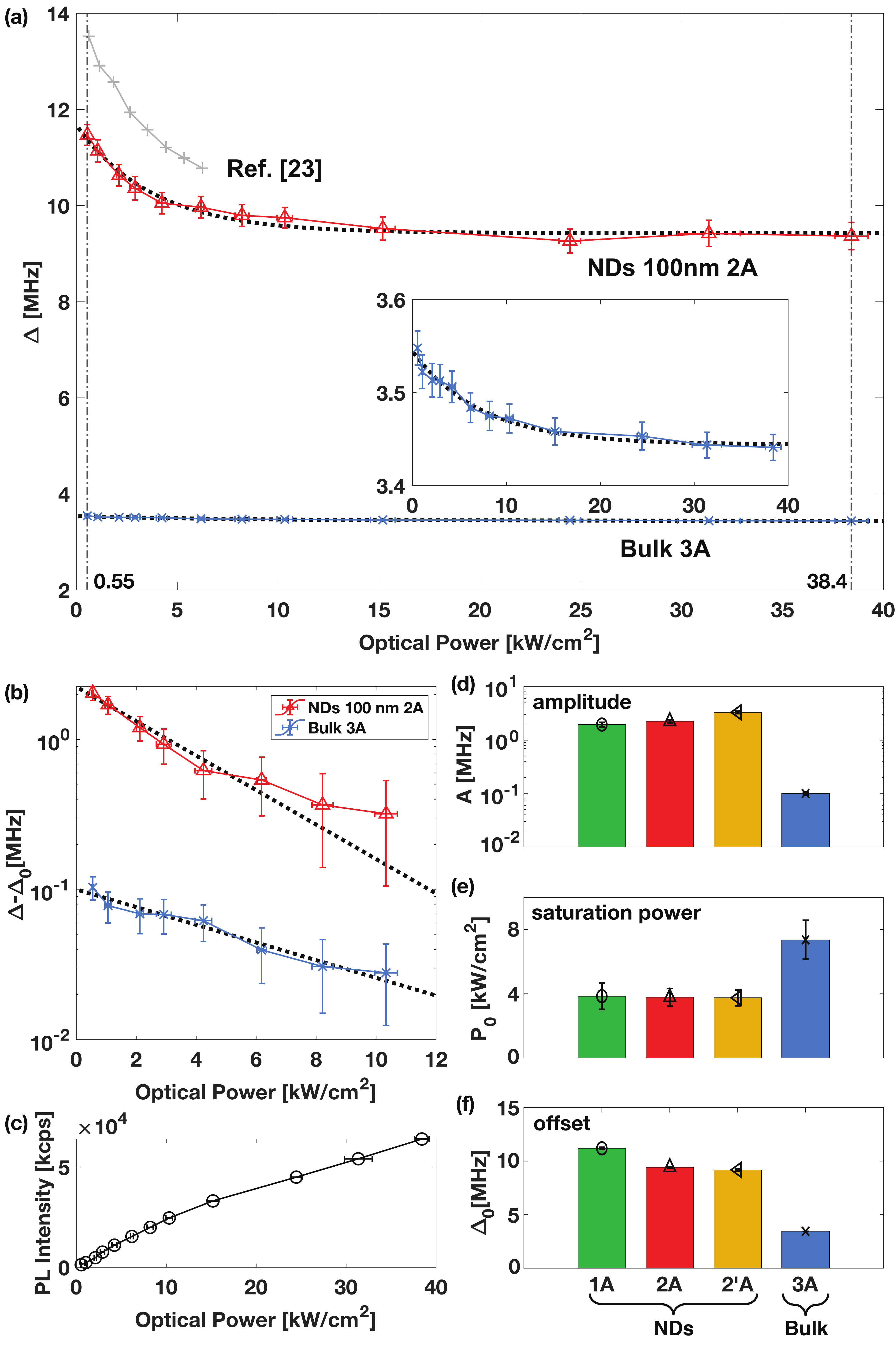}
\caption{\label{fig:4}
(a) 
Behavior of $\Delta$ as a function of $P_\mathrm{opt}$.
The experimental data of 2A and 3A are shown by the markers ($\bigtriangleup$ and $\times$), respectively. For comparison, the data of Ref.~\cite{fujiwara2020} is superposed with the markers ($+$).
The dotted lines result from the fitting to Eq.~(\ref{expfit}).
The inset is a magnified view of the data of the bulk diamond 3A to clarify the exponential decaying behavior.
(b)
Experimental data and the fitted results for 2A and 3A in (a) are presented in the semi-log plot.
The constant offsets $\Delta_0$ have been subtracted from the data.
(c)
PL intensity from NDs in the condition 2A is shown as a function of $P_\mathrm{opt}$.
(d), (e), and (f): 
Fitted parameters (d) $A$, (e) $P_0$, and (f) $\Delta_0$ for the conditions 1A, 2A, 2'A, and 3A [see Table~\ref{tab:condition}] are shown.
Note that the error bars of the data points indicate 1$\sigma$ for all the figures (a)--(f).
}
\end{figure}

Previous study~\cite{fujiwara2020} reported a similar dependence of $\Delta$ on $P_\mathrm{opt}$. 
Their results are superposed in Fig.~4(a) by the markers ($+$).
Significantly, the decaying behavior is almost the same between their results and ours, while they did not reach the optical power to saturate $\Delta$.

It is well established that the PL intensity from an NV center, which is determined by the relaxation rate peculiar to its optical process, saturates for a large $P_\mathrm{opt}$~\cite{dreau2011}. 
However, the present observation is irrelevant as we perform the experiment using a sufficiently small laser intensity such that the PL intensity is linear to $P_\mathrm{opt}$. 
Figure 4(c) confirms that the PL intensity from NDs in the condition 2A is proportional to $P_\mathrm{opt}$.  
Ref.~\cite{fujiwara2020} also treated this sufficiently small optical power region. 
The optical power dependence in such a very small intensity region is unexpected. Our work has quantitatively confirmed Ref.~\cite{fujiwara2020} for a wider optical power region.

It was previously reported~\cite{acosta2013} that the linewidth of the ODMR spectrum of the NV ensemble decreases with increasing $P_\mathrm{opt}$ for an optical power as small as in the present study.
However, they did not mention a decrease in $\Delta$ of the ODMR spectra.
While we observe a systematic change in $\Delta$, no systematic change in the linewidth is detected.

For more quantitative discussion, we analyze the behavior of 2A shown in Fig.~4(a) using  the following exponential fit.
\begin{equation}
\label{expfit}
    \Delta(P_\mathrm{opt}) = A\exp(-P_\mathrm{opt}/P_0)+\Delta_0,
\end{equation}
where $A$, $P_0$, and $\Delta_0$ are the amplitude, the saturation power, and the offset, respectively.
The dotted line in Fig.~4(a) is the result of this fitting.
A semi-log plot of only the first term of Eq.~(\ref{expfit}) is shown in Fig.~4(b) with the same markers as Fig.~4(a).
The linear variation is consistent with the exponential function.
Unlike Fig.~4(a), Fig.~4(b) does not include the previous result~\cite{fujiwara2020} because no convergence value (offset $\Delta_0$) is available.

Then, how about the behavior of the bulk diamond film (the condition 3A)?
Figure~4(a) shows the $P_\mathrm{opt}$ dependence of $\Delta$.
While the decrease of $\Delta$ is not as significant as in NDs (2A), the magnified view in the inset of Fig.~4(a) proves that an exponential decay of $\Delta$ is also present in the bulk diamond case.
Figure~4(b) depicts the decaying component extracted by the fitting to Eq.~(\ref{expfit}), which looks very similar to the 2A case.
The fact suggests a common mechanism behind the present exponential decay of $\Delta$ in the NDs and the bulk diamond, even though different reasons cause the dip splitting.

We find similar behavior in all the measured conditions at zero fields (1A, 2A, 2'A, and 3A in Table I) and obtain the parameters $A$, $P_0$, and $\Delta_0$.
Figure~4(d) shows the obtained amplitude $A$ for the four conditions.
From left to right, the bars indicate the conditions 1A, 2A, 2'A, and 3A, and the vertical axis is expressed on a semi-log scale.
Comparing 1A, 2A, and 2'A, the $A$ values are almost the same for NDs with different grain sizes.
On the other hand, the bulk diamond (3A) has $A$, one order of magnitude smaller than those of NDs (about 1/20).

Figure~4(e) shows the saturation power $P_0$ for different conditions.
While the amplitude $A$ significantly differs between NDs and the bulk diamond, there is relatively little difference in $P_0$ between the two; $P_0 \sim ~3.8~\mathrm{kW/cm^2}$ for NDs and $P_0 \sim 7.4~\mathrm{kW/cm^2}$ for the bulk diamond.
It is vital that the values of $P_0$ are close for different diamonds.
The offsets $\Delta_0$ are shown in Fig.~4(f).
They reduce in the order of conditions 1A, 2A, 2'A, and 3A, which seems to coincide with the degree of deformation of NVs. 
We intuitively expect that the smaller the crystal size is, the greater the deformation tends to be, affecting the sensitivity of the NVs to the optical power. 
We come back to this fact later.

With the results and analysis explained so far, we have established that the ODMR spectra of NVs depend on the excitation light power even when the power is sufficiently small. 
This phenomenon occurs in both NDs and the bulk diamond.
The amplitude of the decay ($A$) largely depends on the samples, but the behavior of exponentially decaying with the optical power characterized by $P_0$ seems an essential feature of NVs. 
The quantitative establishment of the universality of this phenomenon is the main achievement of the present study. 
The fact also means that the excitation light power can be relevant for accurate magnetic field measurements using NVs.

\subsection{Possible Mechanisms}\label{subsec:poss}
We are interested in the possible causes of the observed optical power dependence.
The zero-field splitting (ZFS), the coupling between the NV spin and the magnetic field, and the deformation are the most critical factors in defining the energy structure of an NV center in the ground state~\cite{JelezkoPSS2006}. 
The hyperfine interaction between the NV spin and the neighboring nuclear spins is also often relevant.
Therefore, it is essential as a starting point to investigate whether the present phenomenon is related to these four factors. 
This section will examine them individually and then explore other possibilities.

We start with the ZFS, which might be subject to the optical power through the heating by the laser.
We define the ZFS as the average of the frequencies of the two dips obtained by a double Lorentzian fit.
Around room temperature at zero magnetic fields, the ZFS in the ODMR spectrum decreases linearly with increasing  temperature~\cite{AcostaPRL2010}.
The dependences of ZFS on the optical power in the conditions 1A, 2A, and 3A are shown in Figs.~5(a), (b), and (c), respectively.
The figures indicate no signal of systematic change in ZFS due to the optical power.
Indeed, the variation of ZFS is much smaller than the amplitude $A$ in Fig.~4(d).
Thus, heating by laser irradiation is not responsible for the present optical power dependence.
We estimate the maximum temperature change in this experiment to be about 12~K since the maximum frequency shift observed is approximately 850~kHz, as shown in Fig.~5(a).

\begin{figure} 
\includegraphics[width = \linewidth]{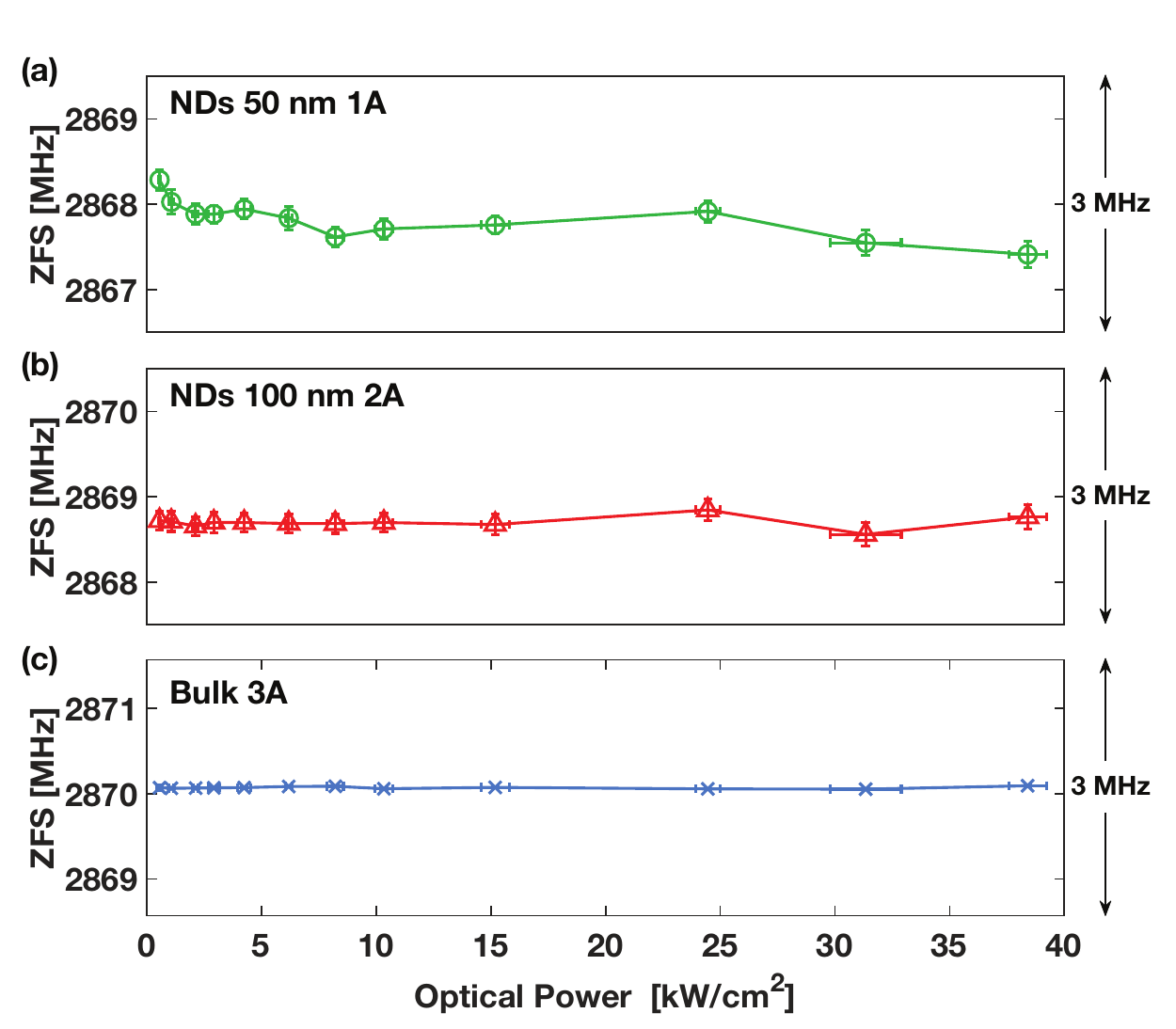}
\caption{\label{fig:5}
Dependence of the ZFS of ODMR spectra for (a) NDs $\phi50$~nm (1A), (b) NDs $\phi100$~nm (2A), and (c) the bulk diamond film (3A) as a function of the optical power $P_\mathrm{opt}$.
The range on the vertical axis is fixed to 3~MHz for (a), (b), and (c).
}
\end{figure}

Next, we discuss the influence of the magnetic field.
The upper and lower panels of Fig.~6 show the ODMR spectra in conditions 2A (zero magnetic field) and 2C (biased magnetic field of $196.7~\mathrm{\mu T}$), respectively [the spectrum shown in the upper is the same as that in the upper panel in Fig.~2(a)].
Both are obtained with the minimum optical power ($P_\mathrm{opt}=0.55~\mathrm{kW/cm^2}$).
The markers are experimental data, and the solid curved lines are fitted by a double Lorentzian function.
The dashed and solid vertical lines show the dip positions obtained by the fit for 2A and 2C, respectively.
As expected from the Zeeman effect, the solid vertical lines are outside the two dashed lines, confirming that $\Delta$ increases in the magnetic field.

\begin{figure} 
\includegraphics[width = \linewidth]{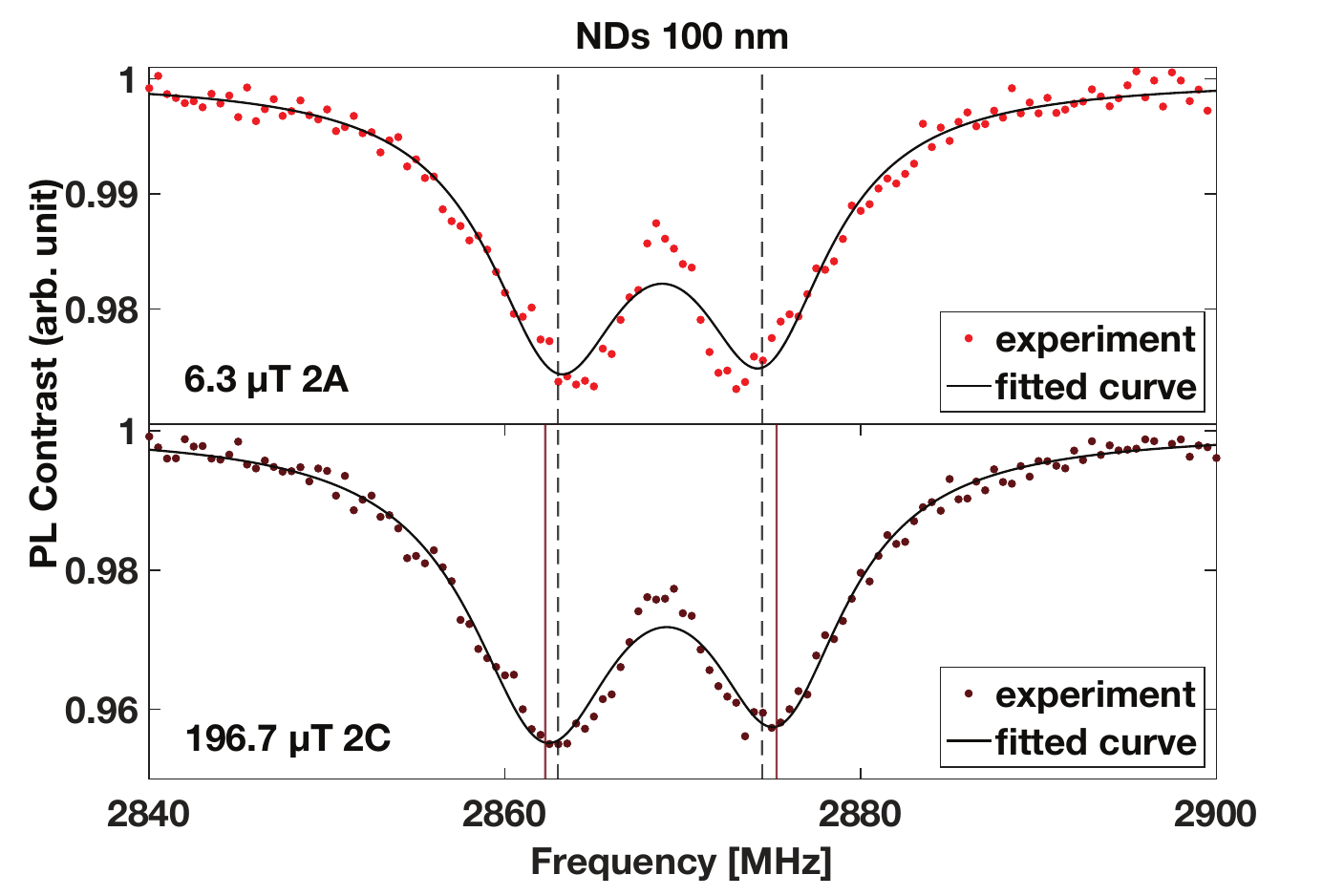}
\caption{\label{fig:6}
ODMR spectra of NDs $\phi100$~nm obtained at $P_\mathrm{opt}=0.55~\mathrm{kW/cm^2}$ in the conditions 2A (the upper panel) and 2C (the lower panel).
The markers are the experimental data, and the solid curved lines result from the fitting.
The dashed and solid vertical lines indicate the fitted dip positions in the upper and lower panels, respectively, supporting that $\Delta$ increases in the magnetic field.
}
\end{figure}

We obtain the spectra for the conditions 2A, 2B, and 2C as $P_\mathrm{opt}$ is modulated.
The acquired behaviors of $\Delta$ are plotted as a function of $P_\mathrm{opt}$ in the inset of Fig.~7(a).
Due to the Zeeman effect, $\Delta$ vertically shifts from 2A to 2B to 2C.
Importantly, there is no significant variation in the spectral shapes of 2A, 2B, and 2C except for this vertical shift.
We obtain the offset $\Delta_0$ by the fitting to Eq.~(\ref{expfit}) and plot $\Delta-\Delta_0$ against $P_\mathrm{opt}$ in the main panel of Fig.~7(a).
The behavior of 2A, 2B, and 2C are superposed on each other almost perfectly.
We plot the amplitude $A$, the saturation power $P_0$, and the offset $\Delta_0$ for each field obtained by the fitting in Figs.~7(b), (c), and (d), respectively.
$\Delta_0$ increases with increasing magnetic field [Fig.~7(d)], reflecting the Zeeman effect, although further quantitative analysis is complicated in this magnetic field region due to the considerable influence of deformation in NDs~\cite{Tsukamoto2022}.
On the other hand, $A$ and $P_0$ do not change significantly as shown in Figs.~7(b) and (c), respectively.
Thus, in our examined regime, there is no visible correlation between the optical power dependence and the magnetic field.

\begin{figure} 
\includegraphics[width = 0.9\linewidth]{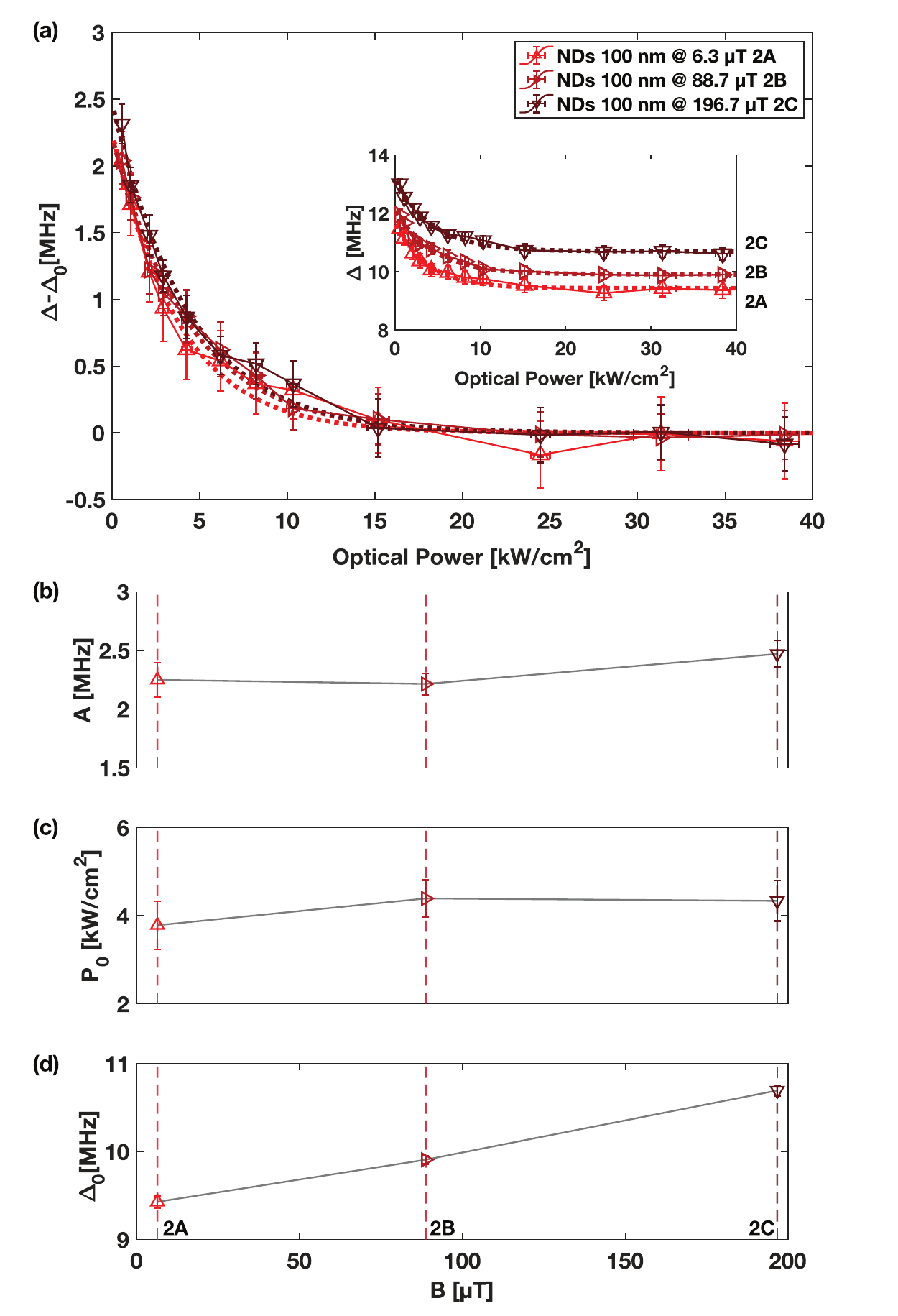}
\caption{\label{fig:7}
(a)
$\Delta-\Delta_0$ as a function of $P_\mathrm{opt}$.
Markers are experimental data for 2A ($\bigtriangleup$), 2B ($\triangleright$), and 2C ($\bigtriangledown$).
The dashed lines result from the fitting to Eq.~(\ref{expfit}).
The inset shows the dependence of $\Delta$.
(b), (c), and (d)
Fitted parameters $A$, $P_0$, and $\Delta_0$ in the different magnetic fields (2A, 2B, and 2C), respectively.
}
\end{figure}

Third, we consider the hyperfine interaction. The optical power dependence in the bulk diamond NVs is minimal, only about 1/20 of that in the nanodiamond NVs [see Figs.~4(a) and 4(d)]. 
However, the contribution of the hyperfine interaction to $\Delta$ is reasonably assumed to be almost similar in the two types of diamonds.
Therefore, if the hyperfine interaction was responsible for the present phenomenon, it would be difficult to explain the marked difference between both. 
Consequently, we can conclude that the hyperfine interaction is not the leading cause of this phenomenon.

As the final factor, we examine the deformation. 
In NDs, the deformation is about 10~MHz [Figs.~2(a) and 4(a)], while the value is well below 1 MHz in the bulk diamond, as discussed in Sec.~IIIB. 
Now, the amplitude $A$ to characterize the optical power dependence is $\sim 2$~MHz for NDs and $\sim 0.1$~MHz for the bulk diamond [Fig.~4(d)]. 
For the former, the ratio of $A$ to the deformation is about $2/10 = 0.2$. 
For the latter, the ratio is at least $0.1/1 = 0.1$ and is comparable to the NDs' case. 
The ratio of ND to bulk diamond deformation also corresponds to the ratio of nitrogen impurity density [see Table~\ref{tab:condition}].
This suggests that either the deformation/impurity itself or the impurity-derived deformation would be responsible for this phenomenon.
Although this argument is not fully quantitative, it suggests a correlation between the deformation/impurity and the optical power dependence.

We infer a reasonable idea of the possible mechanism based on the deformation caused by impurities.
Previous work on single NV centers indicated that the electric field from charge traps causes deformation~\cite{Mittiga2018}.
This might also be the cause with the deformations in the NV ensemble case in our study.
If the charge traps originate from impurities, the magnitude of the deformation will correlate with the impurity density, consistent with our observations.
It is known that the charge state of impurities changes with photoionization.
For example, as the optical power is increased, the time that the NV center retains its charge state decreases exponentially on the millisecond scale~\cite{Aslam2013}.
As this charge generated by photoionization moves around, the electric field would be time-averaged, suppressing deformation.
The relationship between the ionization rate at thermal equilibrium and the photoionization rate determines the coefficient of the exponential change.
When the optical power is sufficiently large, the electric field and crystal strain, which cannot be averaged, remain as a finite deformation.

Ref.~\cite{Mittiga2018} also noted that deformation due to charge can change the shape of the ODMR spectrum to a non-Lorentzian distribution.
This is consistent with the fact that the ODMR spectrum deviates from the double Lorentzian fitting, and its shape changes with optical power [see Figs.~2(a) and (b)].
Investigating both the dip position and its shape will help to elucidate the mechanism.

We note further experimental and theoretical efforts are needed because many parameters could be involved in the mechanism.
On the experimental side, comparing bulk samples with systematically varying impurities and deformations and investigating this optical power-dependent splitting in a single NV center with charge-induced deformation~\cite{Mittiga2018} are helpful.
The magnetic field can be swept over a sufficiently wide range compared to the deformation for bulk samples. This will clarify which parameters of the ground-state Hamiltonian appear to depend on optical power.
Pulsed ODMR~\cite{dreau2011} will provide information on the time the effect of the laser irradiation remains, which can be used to validate the mechanism.
On the theoretical side, it is helpful to investigate what fitting function is appropriate to reproduce the ODMR spectral shape and what defects are candidates for photoionization.

\section{Conclusion}\label{sec:conc}
We investigate the optical power dependence of splitting of the ODMR spectra using various NV ensemble samples.
In addition to reproducing the previous study using NDs~\cite{fujiwara2020}, we find that the optical power dependence saturates in a larger optical power than in their study.
Since we also observe the same phenomenon in the single-crystal diamond, which has very few impurities and non-axisymmetry deformation compared to NDs, we consider our observation due to the NV center's intrinsic nature.
We quantitatively discuss the parameters that could be responsible for this phenomenon and infer that  deformation is an important parameter.
We point out the possible responsibility of slow dynamics in the optical excitation and emission process of single NV centers.

The present optical power dependence can be critical in accurate magnetometry using NVs. 
This effect may degrade the accuracy of the magnetometry using NDs by about a few ten $\mu$T.
Even when using high-quality bulk diamonds, we must be careful when discussing a few $\mu$T magnetic fields around zero magnetic fields.
We can minimize degradation by introducing strong optical power based on the phenomenological exponential behavior discussed here.
Also, we suggest that using diamonds with fewer impurities and deformation can reduce the influence on the accurate magnetic field measurement.
Further experimental verification and theoretical discussion on deformation, impurity densities, and a comprehensive range of magnetic fields will help to identify the mechanism of this phenomenon.

\vspace{0.4cm}
\section*{Acknowledgements}
We thank K. M. Itoh for letting us use the confocal microscope system, and H. Watanabe for his high quality diamond, which we used in the estimation of the spatial resolution of our system [Fig.~1(c)].
We appreciate the fruitful discussion with J. Inoue.
We also thank MEXT-Nanotechnology Platform Program ``Microstructure Analysis Platform" for technical support.
K.S. acknowledges the support of Grants-in-Aid for Scientific Research No. JP22K03524.
K.K. acknowledges the support of Grants-in-Aid for Scientific Research (Nos.~JP23H01103, JP19H00656, and JP19H05826).
T.T. acknowledges the support of MEXT Q-LEAP (JPMXS0118068379), JST CREST (JPMJCR1773), JST Moonshot R\&D (JPMJMS2062), MIC R\&D for construction of a global quantum cryptography network (JPMI00316), JSPS KAKENHI (Nos. JP20H02187 and JP20H05661).

\end{document}